\documentclass[reprint,aps,prb,floats]{revtex4-1}

\usepackage{graphicx}%
\usepackage{dcolumn}%
\usepackage{bm}%

\begin{document}

\title{Mid-Infrared intersubband polaritons in dispersive metal-insulator-metal resonators}
\author{J-M. Manceau$^{1,a)}$, S. Zanotto$^{2,c)}$, T. Ongarello$^{1}$, L. Sorba$^{2}$, A. Tredicucci$^{2,3}$, G. Biasiol$^{4}$, R. Colombelli$^{1,b)}$}
\affiliation{$^1$ Institut d'Electronique Fondamentale, Univ. Paris Sud, UMR8622 CNRS, 91405 Orsay, France\\
$^2$ NEST, Istituto Nanoscienze, CNR and Scuola Normale Superiore, Piazza San Silvestro 12, Pisa, Italy\\
$^3$ Dipartimento di Fisica, Università di Pisa, Largo Pontecorvo 3, I-56127 Pisa, Italy\\
$^4$ Laboratorio TASC, CNR-IOM, Area Science Park, Trieste I-34149, Italy}

\begin{abstract}
We demonstrate room-temperature strong-coupling between a mid-infrared ($\lambda$=9.9 $\mu$m) intersubband transition and the fundamental cavity mode of a metal-insulator-metal resonator. Patterning of the resonator surface enables surface-coupling of the radiation and introduces an energy dispersion which can be probed with angle-resolved reflectivity. In particular, the polaritonic dispersion presents an accessible energy minimum at k=0 where – potentially – polaritons can accumulate. We also show that it is possible to maximize the coupling of photons into the polaritonic states and - simultaneously - to engineer the position of the minimum Rabi splitting at a desired value of the in-plane wavevector. This can be precisely accomplished via a simple post-processing technique. The results are confirmed using the temporal coupled mode theory formalism and their significance in the context of the concept of strong critical coupling is highlighted.
\end{abstract}

\pacs{42.79.Gn, 78.67.De, 85.30.-z}

\maketitle

Light emitting devices based on microcavity polaritons have experienced a tremendous development in the last two decades with the successful demonstration of electroluminescent diodes and optically pumped "bosonic lasers" operating at near-infrared wavelengths \cite{Tsi1,Baj1}. The extension of such devices to mid-infrared (mid-IR) and Terahertz (THz) wavelengths ($\lambda>10\mu$m) has been recently explored, taking advantage of the design flexibility offered by intersubband (ISB) transitions in semiconductor quantum wells.
The strong coupling between an ISB transition (or – more precisely – an ISB plasmon \cite{And1}) and a microcavity photonic mode was first demonstrated in the mid-IR \cite{Din1} and then in the THz range \cite{Tod1}. Devices based on microcavity ISB polaritons hold great potential since in the strong-coupling regime a periodic energy exchange between the light and matter degrees of freedom takes place on an ultrafast time scale (the Rabi oscillation time). On one hand, ISB polaritons can in principle exhibit radiatiave decay times faster than a bare ISB transition. This effect could yield more efficient electroluminescent devices at such wavelengths \cite{Col1,Lib1}. On the other hand, due to their bosonic nature, ISB polaritons are subject to final state stimulation, as it is also the case for  their excitonic counterparts, and they can potentially lead to the demonstration of bosonic mid-IR or THz lasers \cite{Sav1,Kas1,Lib2}, which would not rely on population inversion.
Quantum cascade structures embedded in microcavities have been used to demonstrate electrically pumped light emitting polaritonic devices in the mid-IR \cite{Jou1}. Furthermore, phonon-assisted polariton scattering processes have been observed \cite{Del1}. This constitutes an encouraging step towards the development of efficient electroluminescent polaritonic devices, since it is possible to rely on a proven scattering process. 

However, in the polaritonic light emitting devices (LED) demonstrated to date a key parameter is missing. It is not possible with a total internal reflection cavity geometry to obtain an energy minimum at very low in-plane wavevector ($k_{\parallel}$) values, where the density of states could favor a high bosonic population and – in principle – final state stimulation \cite{Lib2,Lib3}. One dimensional surface plasmon photonic crystal membranes have been successfully used to obtain ISB polaritons at low $k_{\parallel}$, but this geometry is unfortunately incompatible with electrical injection \cite{Zan1}.

In this letter, we demonstrate an optical resonator which offers an energy minimum at $ k_{\parallel}$=0 in the polaritonic dispersion. In essence, we manage to mimic the polaritonic dispersion of exciton-polariton systems based on Fabry-Perot cavities, which has been a crucial tool behind the demonstration of exciton-polariton lasers \cite{Baj1,Car1}. This photonic resonator is compatible with electrical injection. Furthermore, it can be post-processed to maximize the coupling of photons into the polaritonic states and to engineer the position of the minimum energy splitting between upper and lower polaritons at a specific position in k-space.

The device relies on a metal-insulator-metal geometry as depicted in Figure 1a, obtained using standard gold thermo-compression waferbonding technique as detailed, for instance, in Ref. \cite{Xu1}. The bottom mirror is a planar gold layer, 1-$\mu$m-thick. The active region consists of a multiple quantum well (QW) system: 35 repetitions of 8.3-nm-wide GaAs QWs separated by 20-nm-thick Al$_{0.33}$Ga$_{0.67}$As barriers.
\begin{figure}[htp]
\begin{center}
\includegraphics[keepaspectratio=true,scale=1]{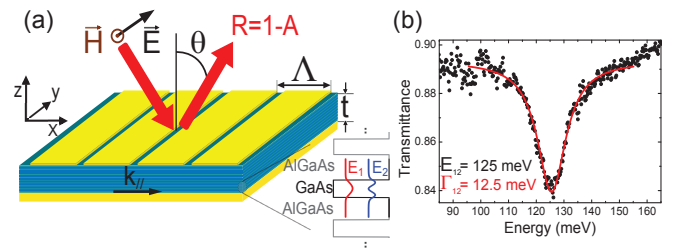}
\caption{(a)Schematic of the device and experimental probing conditions. (b)Quantum well ISB absorption recorded at the Brewster angle and at room temperature. It is centered at 125 meV (E$_{12}$) with a FWHM ($\Gamma_{12}$) of 12.5 meV.}
\vspace{-20pt}
\label{figure1}
\end{center}
\end{figure}
Two of these structures have been grown by molecular beam epitaxy: one with a uniform Si-doping (n$_{2d}$ = $7\times10^{11} cm^{-2}$) within the wells (sample $\#$HM3703), and a second undoped serving as reference (sample $\#$HM3705). The bare ISB transition of the doped sample has been measured in transmission at Brewster angle. A clear absorption peak is detected at an energy of 125 meV ($\lambda$=9.9 $\mu$m, Figure 1b showing a Q-factor of 10. The Q-factor is defined as the ratio E$_{12}$/$\Gamma_{12}$ where $\Gamma_{12}$ is the measured full width at half maximum (FWHM). The surface of the resonator is lithographically patterned with a top 1D metallic grating (Ti/Au, 5/65 nm and a thin layer of Cr, 30nm), with period $\Lambda$ and filling factor ${ff}$, to enable coupling of the system to the external world.
We probe the reflectivity R($\omega$,$\theta$) over a large bandwidth (200 to 2000 $cm^{-1}$), and over a wide angular range ($13^\circ\leq$ $\theta$ $\leq73^\circ$) using a Fourier transform infrared spectrometer equipped with a Globar thermal source. The incoming radiation is P-polarized (electric field in the plane of incidence) using a wire-grid polarizer and the reflected signal is detected with a deuterated triglycine sulfate detector. The absolute reflectivity is obtained by normalizing the sample spectrum against a reference obtained on a planar gold surface (see supplementary material \cite{supp} for reflectance spectra at all the explored angles of incidence). The photonic dispersion (energy vs in-plane wavevector) R(E,$ k_{\parallel}$) is then readily inferred from R($\omega$,$\theta$) using the relationship $k_{\parallel}=\frac{\omega}{c}sin(\theta)$ and E=$\hbar\omega$.
\begin{figure*}[htp]
\begin{center}	\includegraphics[keepaspectratio=true,scale=1.75]{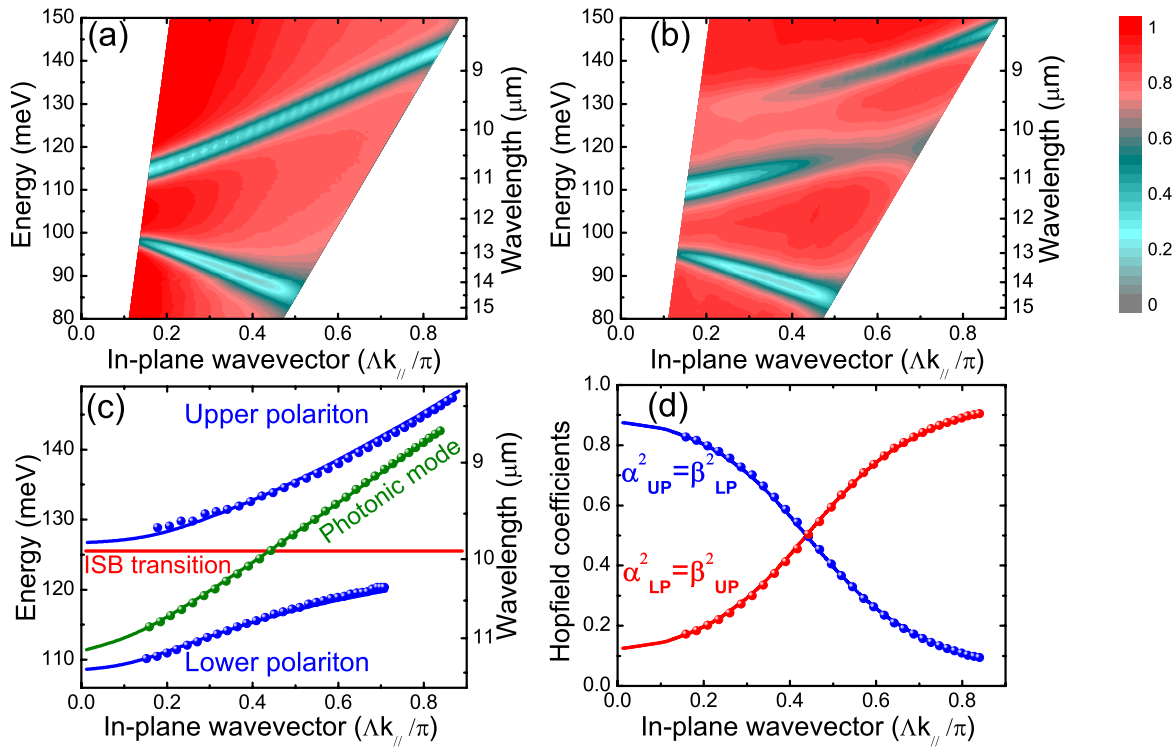}
\caption{(a) Experimental band-diagram of the undoped sample (the color-bar provides the scale for the reflectivity). The grating parameters are $\Lambda$=3.81$\mu$m and ff=83\%. (b) Experimental band-diagram of the doped sample (n$_{2d}$ = $7\times10^{11} cm^{-2}$) with the appearance of the two polariton branches. (c) Dots are the points of minimum reflectivity extracted from the experimental data. Solid lines are the polaritonic dispersions simulated with the RCWA code. (d) Hopfield coefficients deduced from experimental data (dots) and simulated with RCWA (solid lines). The evenly mixed polariton quasi-particles lie at $k_{\parallel}$=0.43.}
\vspace{-20pt}
\label{figure2}
\end{center}
\end{figure*}
The dispersion of the undoped device ($\Lambda$=3.81$\mu$m and ff=83\%) measured at room temperature is shown in Figure 2a and provides information on the resonator. The two dispersive branches of the transverse magnetic (TM) mode folded in the first Brillouin zone are clearly observable. At the $\Gamma$-point (2$^{nd}$ order Bragg scattering) they lie at approximately 110 meV (upper branch) and 100 meV (lower branch), respectively. The upper branch, which is of interest to us since it shows a positive, quadratic dispersion, exhibits a Q-factor of 22 at the energy of the ISB transition (125 meV).\\
We have then probed a doped sample featuring an identical top 1D grating. The sole difference with respect to the previous sample is the presence of the ISB plasmon, which is now active thanks to the presence of electrons from the Si donors. Figure 2b reveals that the upper dispersive branch of the TM mode is now split into two dispersive branches which correspond to the new eigenstates of the system, respectively named lower (LP) and upper (UP) polariton modes. A clear anti-crossing with a minimum splitting of 19 meV has been measured at an incidence angle of 37 degrees. The minimum Rabi splitting, which is to be gauged instead in the (E,$k_{\parallel}$) space, is 17.8 meV and it is measured in this structure at a normalized $k_{\parallel}$= 0.43.\\
The experimental results can be predicted using rigorous coupled wave analysis simulations (RCWA). The numerical details can be found in Refs.\cite{Whi1,Li1}. The ISB	 plasmon couples only to the z-component of the electric field (the component along the growth direction) because of the dipolar selection rule. The multiple QW structure can therefore be modelled as an anisotropic dispersive medium. The in-plane components of the effective permittivity tensor take into account the optical phonons and can be found in Ref.\cite{Pal1}.The ISB transition contribution is included in the z-component of the tensor using the Zaluzny-Nalewajko approach\cite{Zal1}:
\begin{equation}
\epsilon_{z}(\omega)=\epsilon_{\infty}\left(1-f_{0}\frac{\epsilon_{\infty}^{2}}{\epsilon^{2}_{w}}\frac{\omega^{2}_{p}}{\omega^{2}_{12}-\omega^{2}-i\omega\Gamma_{12}}\right)^{-1}
\end{equation}
where $\Gamma_{12}$ is the FWHM of the ISB transition, $\epsilon_{w}$ is the dielectric constant in the well material and $f_{0}$ is the oscillator strength approximated to one for this two level system. The plasma frequency is expressed as follows:
\begin{equation}
\omega_{p}= \sqrt{\frac{\pi e^2 n_{2d}}{\epsilon_{\infty} m^* (L_{b}+L_{w})}}
\end{equation}
where $L_{b}$ and $L_{w}$ are respectively the barrier and well thicknesses, and $n_{2d}$ stands for the dopants’ concentration. Finally the dielectric function of gold is defined according to Ref.\cite{Ord1}.\\
Figure 2c shows the simulated polariton dispersion (continuous lines) superimposed onto the reflectivity minima as extracted from the experimental data (dots). Theory and experiment are in excellent agreement and - most importantly – the dispersion presents an energy minimum at $k_{\parallel}$=0. The eigenstates of the coupled system are a linear superposition of the uncoupled eigenstates $|\psi_{i,k}\rangle$  (with $|\psi_{1,k}\rangle$ the fundamental state and $|\psi_{2,k}\rangle$ the excited state of the ISB plasmon) and $|n\rangle$ the state with n photons in the cavity, as follows:
\begin{equation}
•|UP\rangle=\alpha_{UP}|\psi_{1,k},1\rangle+\beta_{UP}|\psi_{2,k},0\rangle,
 \end{equation}
\begin{equation}
•|LP\rangle=\alpha_{LP}|\psi_{1,k},1\rangle+\beta_{LP}|\psi_{2,k},0\rangle
 \end{equation}
where UP is the higher energy eigenstate and LP is the lower energy eigenstate. The coefficients $\alpha_{LP,UP}$, $\beta_{LP,UP}$ are called the Hopfield coefficients \cite{Hop1} and they permit to gauge the weight of the photonic/material components within the polariton branches. Using analytical formulas (See supplementary material \cite{supp} for details on the extraction of the Hopfield coefficients) and the set of experimental and simulated data, we have inferred the fractional contribution of each component, as reported in Fig. 2d. An even mixing (50-50\%) is observed at a normalized $k_{\parallel}$ value of 0.43, which corresponds to the minimum Rabi splitting. Further away from the minimum splitting, the quasi-particles lose their mixed nature. Note however that even at $k_{\parallel}$ =0 the LP still maintains more than 10\% of photonic component.\\
In the perspective of developing polaritonic light emitting devices and especially lasers, a few aspects need to be taken into account. First, the control of the Hopfield coefficients is of crucial importance, since it permits to tailor / maximise the polaritonic lifetime. For instance, the possibility to engineer the minimum energy splitting at a desired position in the $k_{\parallel}$-space would be an important asset. The second key point is the need to efficiently populate the polaritonic states, via optical or electrical pumping. The first point, and also the second in the case of optical pumping, can be addressed with a post-processing approach that we have recently demonstrated on a similar resonator geometry \cite{Man7}. In the general framework of temporal coupled mode theory (TCMT) \cite{Hau1,Fan1}, the device employing the undoped active region can be described as a one port system. Its electromagnetic coupling to the external world is described by a radiative damping rate $\gamma_{r}$, while the resonator losses are lumped into a non-radiative rate $\gamma_{nr}$. When the two damping rates are matched ($\gamma_{r}=\gamma_{nr}$), all the incoming photons are coupled into the resonant photonic mode, a situation known as critical coupling \cite{Man7,Cai1}.\\
Interestingly, this general concept also holds for the regime of strong coupling, as developed in Ref. \cite{Zan2}. To this scope, it is necessary to add an additional oscillator to model the ISB plasmon, and introduce the light-matter coupling constant \cite{Auf1}. The TCMT equations become:
\begin{equation}
 \frac{db}{dt}=(i\omega_{12}-\gamma_{12})b+i\Omega a
\end{equation}		
\begin{equation}
 \frac{da}{dt}=(i\omega_{c}-\gamma_{c})b+i\Omega b+ks^+
 \end{equation}
\begin{equation}
 s^-=cs^+ +ad
\end{equation}					
where $|s^\pm|^2$ is the energy flux per unit time ingoing (outgoing) into (out of) the system; $\vert a\vert^2$ is the total electromagnetic energy stored in the cavity; $\vert b\vert^2$ is the total energy stored in the matter resonator; $\gamma_{12}$ is the ISB plasmon damping rate and $\omega_{12}$ its frequency; $\gamma_{c}= \gamma_{r}+\gamma_{nr}$ is the total cavity damping rate; and $\Omega$ is the light-matter coupling constant (i.e. the Rabi frequency). If $2\Omega\gg\mid\gamma_{r}+ \gamma_{nr}-\gamma_{12}\mid^2$ the system is in the strong coupling regime [32]. It is possible to show that perfect energy feeding into the polaritonic states takes place under a novel strong critical coupling (SCC) condition which is $\gamma_{r}=\gamma_{nr} +\gamma_{12}$. The TCMT description of the system is of crucial importance as it allows one to naturally describe on the same footing the coupling between the resonator and the material excitations – which gives rise to polaritons – simultaneously with the coupling of the system to the external world. Note: the damping rates used in the frame of TCMT are twice smaller than the experimental ones since the time-dependent amplitude is now considered.
\begin{figure}[htp]
\begin{center}
\includegraphics[keepaspectratio=true,scale=1]{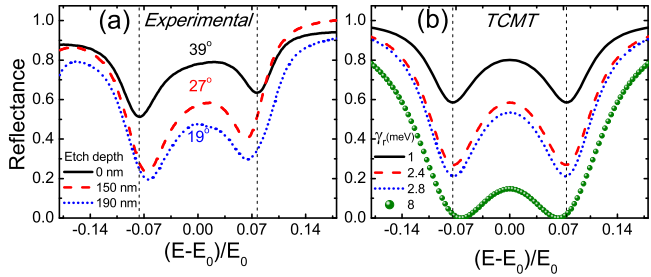}
\caption{(a) Angularly-resolved experimental reflectance at the minimum energy splitting for different etching depths. The angle of measurement is mentioned next to the curve. (b) Simulated reflectance with the CMT. The radiative damping rate ($\gamma_{r}$ in meV) is the tuning parameter. The round points represent the regime of strong critical coupling.}
\vspace{-10pt}
\label{figure3}
\end{center}
\end{figure}
To experimentally demonstrate the existence of this regime, we iteratively etched the semiconductor material between the grating metallic fingers with an inductively coupled plasma reactor and using the metallic grating as mask. Etching has a direct impact on $\gamma_{r}$ while it essentially does not affect $\gamma_{nr}$ as demonstrated in Ref. [28]. After each etching step, we measure the polaritonic dispersion with our experimental set-up. Figure 3a shows the reflectivity of the sample at an incidence angle corresponding to the minimum splitting. The energy scale (x-axis) is renormalized by the frequency of the ISB transition. As the sample is etched, more photons get coupled into the polaritonic states. From an initial absorption of $\sim50\%$, we reach $80\%$ at an etch depth of 190 nm. Further etching brings the minimum splitting point outside the experimentally accessible light cone and makes the measurement impossible. This behaviour is well reproduced by the TCMT, as shown by the simulations in Fig. 3b, which report $|r(\omega)|^2=|{s^-}/{s^+}|^2$ as obtained from the equations above. The sole fitting parameter is $\gamma_{r}$, which reveals that the resonator is under-coupled: the etching procedure increases $\gamma_{r}$ thus driving it towards the critical coupling condition. Note: the slight reduction of the splitting, as inferred from the reflectance minima in Fig. 3a, further confirms that we are affecting the damping rates of the system.\\
We can now explore how to engineer the position of the minimum Rabi splitting with respect to the in-plane wave-vector $k_{\parallel}$. As the sample gets etched, the effective refractive index contrast increases as we are carving air-holes into a 1D photonic crystal. This gradual change of index contrast broadens the gap at $k_{\parallel}$=0 and permits to spectrally blue-shift the upper photonic branch of the resonator to a desired value. Figure 4a presents the Hopfield coefficients extracted from the experimental results on the non-etched sample (black line) and after a 290-nm-deep etch (dotted line). The Hopfield coefficients of this last etched configuration have been deduced with the RCWA code (See supplementary material \cite{supp} for details on the extraction of the Hopfield coefficients). In this case, we have been able to engineer the regime of evenly mixed polaritonic states (i.e. equal Hopfield coefficients) from $k_{\parallel}$=0.43 to $k_{\parallel}$=0. This post-processing approach which allows one to tune the position of the minimum splitting is even more remarkable as it occurs simultaneously with the onset of the strong critical coupling regime.
\begin{figure}[htp]
\begin{center}
\includegraphics[keepaspectratio=true,scale=0.75]{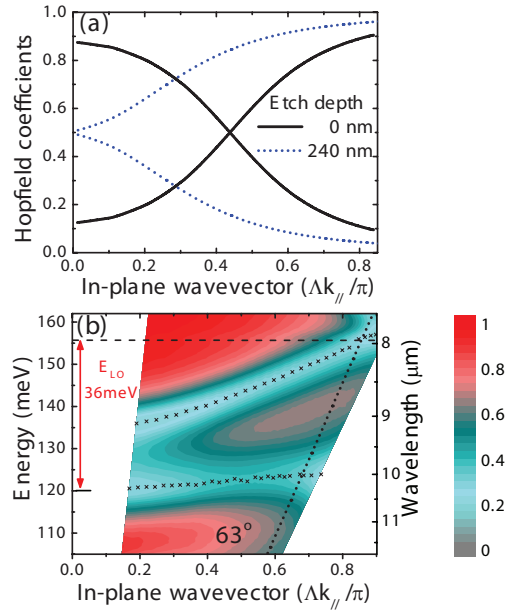}
\caption{(a) Calculated Hopfield coefficients for the initial device and after 290 nm of etching (dots). The evenly mixed polaritons occur now at the $\Gamma$ point of the polaritonic dispersion. (b) Experimental band-diagram (the color-bar provides the scale for the reflectivity) of the device final configuration with the injection angle for optical pumping experiment at 55$^\circ$(dotted line) and the representation of a longitudinal optical phonon energy separation (E$_{LO}$) as a potential scattering mechanism. The crosses represent the reflectance minima.}
\vspace{-10pt}
\label{figure4}
\end{center}
\end{figure}
%
The final resonator configuration (Fig. 4b) is of particular relevance in the context of polariton scattering experiments under optical or electrical pumping. Figure 4b presents the polaritonic dispersion at the final etch depth of 290 nm: the minimum splitting occurs at $k_{\parallel}$=0 and the light coupling is greatly improved over the whole dispersion diagram. In particular, it is possible to envision an optical pumping experiment where an incident beam at $\theta=63^\circ$ efficiently generates ISB polaritons, since the absorption is larger than $80\%$. Efficient population of the $k_{\parallel}$=0 state in turn relies on polariton-LO phonon scattering [17]. On one hand, this is a promising architecture for polaritonic LEDs. On the other hand, with the appropriate lifetime engineering, the occupation number of the ground state could be improved and even reach the value of 1, necessary for the final state stimulation process [16].\\
In conclusion, we have demonstrated room temperature strong coupling between a mid-infrared intersubband plasmon and the fundamental cavity mode of a metal-insulator-metal resonator, with a polaritonic dispersion presenting an energy minimum at $k_{\parallel}$=0. We have also shown the possibility to maximize the coupling of photons into the polaritonic states (close to critical coupling) and simultaneously to engineer the position of the minimum Rabi splitting at a desired value of the in-plane wavevector. These results constitute a building block for future developments of long wavelength optically or electrically pumped polaritonic light emitting devices.
\acknowledgments 
We thank N. Isac for help with the wafer-bonding process, and we acknowledge financial support from the Triangle de la Physique (Project 'INTENSE'), and from the French National Research Agency (ANR-09-NANO-017 'HI-TEQ'). This work was partly supported by the French RENATECH  network. R.C. acknowledges partial support from the ERC 'GEM' grant (Grant agreement $\#$306661), and A.T.  from the ERC Advanced Grant 'SoulMan'.


\begin{thebibliography}{28}%
\makeatletter
\providecommand \@ifxundefined [1]{%
 \@ifx{#1\undefined}
}%
\providecommand \@ifnum [1]{%
 \ifnum #1\expandafter \@firstoftwo
 \else \expandafter \@secondoftwo
 \fi
}%
\providecommand \@ifx [1]{%
 \ifx #1\expandafter \@firstoftwo
 \else \expandafter \@secondoftwo
 \fi
}%
\providecommand \natexlab [1]{#1}%
\providecommand \enquote  [1]{``#1''}%
\providecommand \bibnamefont  [1]{#1}%
\providecommand \bibfnamefont [1]{#1}%
\providecommand \citenamefont [1]{#1}%
\providecommand \href@noop [0]{\@secondoftwo}%
\providecommand \href [0]{\begingroup \@sanitize@url \@href}%
\providecommand \@href[1]{\@@startlink{#1}\@@href}%
\providecommand \@@href[1]{\endgroup#1\@@endlink}%
\providecommand \@sanitize@url [0]{\catcode `\\12\catcode `\$12\catcode
  `\&12\catcode `\#12\catcode `\^12\catcode `\_12\catcode `\%12\relax}%
\providecommand \@@startlink[1]{}%
\providecommand \@@endlink[0]{}%
\providecommand \url  [0]{\begingroup\@sanitize@url \@url }%
\providecommand \@url [1]{\endgroup\@href {#1}{\urlprefix }}%
\providecommand \urlprefix  [0]{URL }%
\providecommand \Eprint [0]{\href }%
\providecommand \doibase [0]{http://dx.doi.org/}%
\providecommand \selectlanguage [0]{\@gobble}%
\providecommand \bibinfo  [0]{\@secondoftwo}%
\providecommand \bibfield  [0]{\@secondoftwo}%
\providecommand \translation [1]{[#1]}%
\providecommand \BibitemOpen [0]{}%
\providecommand \bibitemStop [0]{}%
\providecommand \bibitemNoStop [0]{.\EOS\space}%
\providecommand \EOS [0]{\spacefactor3000\relax}%
\providecommand \BibitemShut  [1]{\csname bibitem#1\endcsname}%
\let\auto@bib@innerbib\@empty
\item[$^{a)}$] Email: jean-michel.manceau@u-psud.fr
\item[$^{b)}$] Email: raffaele.colombelli@u-psud.fr
\item[$^{c)}$] Present address: Dipartimento di Elettronica, Informazione e Bioingegneria, Politecnico di Milano, Via Colombo 81, 20133 Milano, Italy
\vskip 5pt
\bibitem [{\citenamefont {Tsintzos}\ \emph {et~al.}(2008)\citenamefont
  {Tsintzos}, \citenamefont {Pelekanos}, \citenamefont {Konstantinidis},
  \citenamefont {Hatzopoulos},\ and\ \citenamefont {Savvidis}}]{Tsi1}%
  \BibitemOpen
  \bibfield  {author} {\bibinfo {author} {\bibfnamefont {S.~I.}\ \bibnamefont
  {Tsintzos}}, \bibinfo {author} {\bibfnamefont {N.~T.}\ \bibnamefont
  {Pelekanos}}, \bibinfo {author} {\bibfnamefont {G.}~\bibnamefont
  {Konstantinidis}}, \bibinfo {author} {\bibfnamefont {Z.}~\bibnamefont
  {Hatzopoulos}}, \ and\ \bibinfo {author} {\bibfnamefont {P.~G.}\ \bibnamefont
  {Savvidis}},\ }\href@noop {} {\bibfield  {journal} {\bibinfo  {journal}
  {Nature}\ }\textbf {\bibinfo {volume} {453}},\ \bibinfo {pages} {372}
  (\bibinfo {year} {2008})}\BibitemShut {NoStop}%
\bibitem [{\citenamefont {Bajoni}(2012)}]{Baj1}%
  \BibitemOpen
  \bibfield  {author} {\bibinfo {author} {\bibfnamefont {D.}~\bibnamefont
  {Bajoni}},\ }\href@noop {} {\bibfield  {journal} {\bibinfo  {journal}
  {Journal of Physics D: Applied Physics}\ }\textbf {\bibinfo {volume} {45}},\
  \bibinfo {pages} {313001} (\bibinfo {year} {2012})}\BibitemShut {NoStop}%
\bibitem [{\citenamefont {Ando}, \citenamefont {Fowler},\ and\ \citenamefont
  {Stern}(1982)}]{And1}%
  \BibitemOpen
  \bibfield  {author} {\bibinfo {author} {\bibfnamefont {T.}~\bibnamefont
  {Ando}}, \bibinfo {author} {\bibfnamefont {A.~B.}\ \bibnamefont {Fowler}}, \
  and\ \bibinfo {author} {\bibfnamefont {F.}~\bibnamefont {Stern}},\
  }\href@noop {} {\bibfield  {journal} {\bibinfo  {journal} {Reviews of Modern
  Physics}\ }\textbf {\bibinfo {volume} {54}},\ \bibinfo {pages} {437}
  (\bibinfo {year} {1982})}\BibitemShut {NoStop}%
\bibitem [{\citenamefont {Dini}\ \emph {et~al.}(2003)\citenamefont {Dini},
  \citenamefont {Köhler}, \citenamefont {Tredicucci}, \citenamefont
  {Biasiol},\ and\ \citenamefont {Sorba}}]{Din1}%
  \BibitemOpen
  \bibfield  {author} {\bibinfo {author} {\bibfnamefont {D.}~\bibnamefont
  {Dini}}, \bibinfo {author} {\bibfnamefont {R.}~\bibnamefont {Köhler}},
  \bibinfo {author} {\bibfnamefont {A.}~\bibnamefont {Tredicucci}}, \bibinfo
  {author} {\bibfnamefont {G.}~\bibnamefont {Biasiol}}, \ and\ \bibinfo
  {author} {\bibfnamefont {L.}~\bibnamefont {Sorba}},\ }\href@noop {}
  {\bibfield  {journal} {\bibinfo  {journal} {Physical Review Letters}\
  }\textbf {\bibinfo {volume} {90}},\ \bibinfo {pages} {116401} (\bibinfo
  {year} {2003})}\BibitemShut {NoStop}%
\bibitem [{\citenamefont {Todorov}\ \emph {et~al.}(2009)\citenamefont
  {Todorov}, \citenamefont {Andrews}, \citenamefont {Sagnes}, \citenamefont
  {Colombelli}, \citenamefont {Klang}, \citenamefont {Strasser},\ and\
  \citenamefont {Sirtori}}]{Tod1}%
  \BibitemOpen
  \bibfield  {author} {\bibinfo {author} {\bibfnamefont {Y.}~\bibnamefont
  {Todorov}}, \bibinfo {author} {\bibfnamefont {A.~M.}\ \bibnamefont
  {Andrews}}, \bibinfo {author} {\bibfnamefont {I.}~\bibnamefont {Sagnes}},
  \bibinfo {author} {\bibfnamefont {R.}~\bibnamefont {Colombelli}}, \bibinfo
  {author} {\bibfnamefont {P.}~\bibnamefont {Klang}}, \bibinfo {author}
  {\bibfnamefont {G.}~\bibnamefont {Strasser}}, \ and\ \bibinfo {author}
  {\bibfnamefont {C.}~\bibnamefont {Sirtori}},\ }\href@noop {} {\bibfield
  {journal} {\bibinfo  {journal} {Physical Review Letters}\ }\textbf {\bibinfo
  {volume} {102}},\ \bibinfo {pages} {186402} (\bibinfo {year}
  {2009})}\BibitemShut {NoStop}%
\bibitem [{\citenamefont {Colombelli}\ \emph {et~al.}(2005)\citenamefont
  {Colombelli}, \citenamefont {Ciuti}, \citenamefont {Chassagneux},\ and\
  \citenamefont {Sirtori}}]{Col1}%
  \BibitemOpen
  \bibfield  {author} {\bibinfo {author} {\bibfnamefont {R.}~\bibnamefont
  {Colombelli}}, \bibinfo {author} {\bibfnamefont {C.}~\bibnamefont {Ciuti}},
  \bibinfo {author} {\bibfnamefont {Y.}~\bibnamefont {Chassagneux}}, \ and\
  \bibinfo {author} {\bibfnamefont {C.}~\bibnamefont {Sirtori}},\ }\href@noop
  {} {\bibfield  {journal} {\bibinfo  {journal} {Semiconductor Science and
  Technology}\ }\textbf {\bibinfo {volume} {20}},\ \bibinfo {pages} {985}
  (\bibinfo {year} {2005})}\BibitemShut {NoStop}%
\bibitem [{\citenamefont {De~Liberato}\ and\ \citenamefont
  {Ciuti}(2008)}]{Lib1}%
  \BibitemOpen
  \bibfield  {author} {\bibinfo {author} {\bibfnamefont {S.}~\bibnamefont
  {De~Liberato}}\ and\ \bibinfo {author} {\bibfnamefont {C.}~\bibnamefont
  {Ciuti}},\ }\href@noop {} {\bibfield  {journal} {\bibinfo  {journal}
  {Physical Review B}\ }\textbf {\bibinfo {volume} {77}},\ \bibinfo {pages}
  {155321} (\bibinfo {year} {2008})}\BibitemShut {NoStop}%
\bibitem [{\citenamefont {Savvidis}\ \emph {et~al.}(2000)\citenamefont
  {Savvidis}, \citenamefont {Baumberg}, \citenamefont {Stevenson},
  \citenamefont {Skolnick}, \citenamefont {Whittaker},\ and\ \citenamefont
  {Roberts}}]{Sav1}%
  \BibitemOpen
  \bibfield  {author} {\bibinfo {author} {\bibfnamefont {P.~G.}\ \bibnamefont
  {Savvidis}}, \bibinfo {author} {\bibfnamefont {J.~J.}\ \bibnamefont
  {Baumberg}}, \bibinfo {author} {\bibfnamefont {R.~M.}\ \bibnamefont
  {Stevenson}}, \bibinfo {author} {\bibfnamefont {M.~S.}\ \bibnamefont
  {Skolnick}}, \bibinfo {author} {\bibfnamefont {D.~M.}\ \bibnamefont
  {Whittaker}}, \ and\ \bibinfo {author} {\bibfnamefont {J.~S.}\ \bibnamefont
  {Roberts}},\ }\href@noop {} {\bibfield  {journal} {\bibinfo  {journal}
  {Physical Review Letters}\ }\textbf {\bibinfo {volume} {84}},\ \bibinfo
  {pages} {1547} (\bibinfo {year} {2000})}\BibitemShut {NoStop}%
\bibitem [{\citenamefont {Kasprzak}\ \emph {et~al.}(2006)\citenamefont
  {Kasprzak}, \citenamefont {Richard}, \citenamefont {Kundermann},
  \citenamefont {Baas}, \citenamefont {Jeambrun}, \citenamefont {Keeling},
  \citenamefont {Marchetti}, \citenamefont {Szymanska}, \citenamefont {Andre},
  \citenamefont {Staehli}, \citenamefont {Savona}, \citenamefont {Littlewood},
  \citenamefont {Deveaud},\ and\ \citenamefont {Dang}}]{Kas1}%
  \BibitemOpen
  \bibfield  {author} {\bibinfo {author} {\bibfnamefont {J.}~\bibnamefont
  {Kasprzak}}, \bibinfo {author} {\bibfnamefont {M.}~\bibnamefont {Richard}},
  \bibinfo {author} {\bibfnamefont {S.}~\bibnamefont {Kundermann}}, \bibinfo
  {author} {\bibfnamefont {A.}~\bibnamefont {Baas}}, \bibinfo {author}
  {\bibfnamefont {P.}~\bibnamefont {Jeambrun}}, \bibinfo {author}
  {\bibfnamefont {J.~M.~J.}\ \bibnamefont {Keeling}}, \bibinfo {author}
  {\bibfnamefont {F.~M.}\ \bibnamefont {Marchetti}}, \bibinfo {author}
  {\bibfnamefont {M.~H.}\ \bibnamefont {Szymanska}}, \bibinfo {author}
  {\bibfnamefont {R.}~\bibnamefont {Andre}}, \bibinfo {author} {\bibfnamefont
  {J.~L.}\ \bibnamefont {Staehli}}, \bibinfo {author} {\bibfnamefont
  {V.}~\bibnamefont {Savona}}, \bibinfo {author} {\bibfnamefont {P.~B.}\
  \bibnamefont {Littlewood}}, \bibinfo {author} {\bibfnamefont
  {B.}~\bibnamefont {Deveaud}}, \ and\ \bibinfo {author} {\bibfnamefont
  {L.~S.}\ \bibnamefont {Dang}},\ }\href@noop {} {\bibfield  {journal}
  {\bibinfo  {journal} {Nature}\ }\textbf {\bibinfo {volume} {443}},\ \bibinfo
  {pages} {409} (\bibinfo {year} {2006})}\BibitemShut {NoStop}%
\bibitem [{\citenamefont {De~Liberato}\ and\ \citenamefont
  {Ciuti}(2009)}]{Lib2}%
  \BibitemOpen
  \bibfield  {author} {\bibinfo {author} {\bibfnamefont {S.}~\bibnamefont
  {De~Liberato}}\ and\ \bibinfo {author} {\bibfnamefont {C.}~\bibnamefont
  {Ciuti}},\ }\href@noop {} {\bibfield  {journal} {\bibinfo  {journal}
  {Physical Review Letters}\ }\textbf {\bibinfo {volume} {102}},\ \bibinfo
  {pages} {136403} (\bibinfo {year} {2009})}\BibitemShut {NoStop}%
\bibitem [{\citenamefont {Jouy}\ \emph {et~al.}(2010)\citenamefont {Jouy},
  \citenamefont {Vasanelli}, \citenamefont {Todorov}, \citenamefont {Sapienza},
  \citenamefont {Colombelli}, \citenamefont {Gennser},\ and\ \citenamefont
  {Sirtori}}]{Jou1}%
  \BibitemOpen
  \bibfield  {author} {\bibinfo {author} {\bibfnamefont {P.}~\bibnamefont
  {Jouy}}, \bibinfo {author} {\bibfnamefont {A.}~\bibnamefont {Vasanelli}},
  \bibinfo {author} {\bibfnamefont {Y.}~\bibnamefont {Todorov}}, \bibinfo
  {author} {\bibfnamefont {L.}~\bibnamefont {Sapienza}}, \bibinfo {author}
  {\bibfnamefont {R.}~\bibnamefont {Colombelli}}, \bibinfo {author}
  {\bibfnamefont {U.}~\bibnamefont {Gennser}}, \ and\ \bibinfo {author}
  {\bibfnamefont {C.}~\bibnamefont {Sirtori}},\ }\href@noop {} {\bibfield
  {journal} {\bibinfo  {journal} {Physical Review B}\ }\textbf {\bibinfo
  {volume} {82}} (\bibinfo {year} {2010})}\BibitemShut {NoStop}%
\bibitem [{\citenamefont {Delteil}\ \emph {et~al.}(2011)\citenamefont
  {Delteil}, \citenamefont {Vasanelli}, \citenamefont {Jouy}, \citenamefont
  {Barate}, \citenamefont {Moreno}, \citenamefont {Teissier}, \citenamefont
  {Baranov},\ and\ \citenamefont {Sirtori}}]{Del1}%
  \BibitemOpen
  \bibfield  {author} {\bibinfo {author} {\bibfnamefont {A.}~\bibnamefont
  {Delteil}}, \bibinfo {author} {\bibfnamefont {A.}~\bibnamefont {Vasanelli}},
  \bibinfo {author} {\bibfnamefont {P.}~\bibnamefont {Jouy}}, \bibinfo {author}
  {\bibfnamefont {D.}~\bibnamefont {Barate}}, \bibinfo {author} {\bibfnamefont
  {J.~C.}\ \bibnamefont {Moreno}}, \bibinfo {author} {\bibfnamefont
  {R.}~\bibnamefont {Teissier}}, \bibinfo {author} {\bibfnamefont {A.~N.}\
  \bibnamefont {Baranov}}, \ and\ \bibinfo {author} {\bibfnamefont
  {C.}~\bibnamefont {Sirtori}},\ }\href@noop {} {\bibfield  {journal} {\bibinfo
   {journal} {Physical Review B}\ }\textbf {\bibinfo {volume} {83}},\ \bibinfo
  {pages} {081404} (\bibinfo {year} {2011})}\BibitemShut {NoStop}%
\bibitem [{\citenamefont {De~Liberato}, \citenamefont {Ciuti},\ and\
  \citenamefont {Phillips}(2013)}]{Lib3}%
  \BibitemOpen
  \bibfield  {author} {\bibinfo {author} {\bibfnamefont {S.}~\bibnamefont
  {De~Liberato}}, \bibinfo {author} {\bibfnamefont {C.}~\bibnamefont {Ciuti}},
  \ and\ \bibinfo {author} {\bibfnamefont {C.~C.}\ \bibnamefont {Phillips}},\
  }\href@noop {} {\bibfield  {journal} {\bibinfo  {journal} {Physical Review
  B}\ }\textbf {\bibinfo {volume} {87}} (\bibinfo {year} {2013})}\BibitemShut
  {NoStop}%
\bibitem [{\citenamefont {Zanotto}\ \emph {et~al.}(2012)\citenamefont
  {Zanotto}, \citenamefont {Degl'Innocenti}, \citenamefont {Xu}, \citenamefont
  {Sorba}, \citenamefont {Tredicucci},\ and\ \citenamefont {Biasiol}}]{Zan1}%
  \BibitemOpen
  \bibfield  {author} {\bibinfo {author} {\bibfnamefont {S.}~\bibnamefont
  {Zanotto}}, \bibinfo {author} {\bibfnamefont {R.}~\bibnamefont
  {Degl'Innocenti}}, \bibinfo {author} {\bibfnamefont {J.-H.}\ \bibnamefont
  {Xu}}, \bibinfo {author} {\bibfnamefont {L.}~\bibnamefont {Sorba}}, \bibinfo
  {author} {\bibfnamefont {A.}~\bibnamefont {Tredicucci}}, \ and\ \bibinfo
  {author} {\bibfnamefont {G.}~\bibnamefont {Biasiol}},\ }\href@noop {}
  {\bibfield  {journal} {\bibinfo  {journal} {Physical Review B}\ }\textbf
  {\bibinfo {volume} {86}},\ \bibinfo {pages} {201302} (\bibinfo {year}
  {2012})}\BibitemShut {NoStop}%
\bibitem [{\citenamefont {Carusotto}\ and\ \citenamefont {Ciuti}(2013)}]{Car1}%
  \BibitemOpen
  \bibfield  {author} {\bibinfo {author} {\bibfnamefont {I.}~\bibnamefont
  {Carusotto}}\ and\ \bibinfo {author} {\bibfnamefont {C.}~\bibnamefont
  {Ciuti}},\ }\href@noop {} {\bibfield  {journal} {\bibinfo  {journal} {Reviews
  of Modern Physics}\ }\textbf {\bibinfo {volume} {85}},\ \bibinfo {pages}
  {299} (\bibinfo {year} {2013})}\BibitemShut {NoStop}%
\bibitem [{\citenamefont {Xu}\ \emph {et~al.}(2012)\citenamefont {Xu},
  \citenamefont {Colombelli}, \citenamefont {Khanna}, \citenamefont
  {Belarouci}, \citenamefont {Letartre}, \citenamefont {Li}, \citenamefont
  {Linfield}, \citenamefont {Davies}, \citenamefont {Beere},\ and\
  \citenamefont {Ritchie}}]{Xu1}%
  \BibitemOpen
  \bibfield  {author} {\bibinfo {author} {\bibfnamefont {G.}~\bibnamefont
  {Xu}}, \bibinfo {author} {\bibfnamefont {R.}~\bibnamefont {Colombelli}},
  \bibinfo {author} {\bibfnamefont {S.~P.}\ \bibnamefont {Khanna}}, \bibinfo
  {author} {\bibfnamefont {A.}~\bibnamefont {Belarouci}}, \bibinfo {author}
  {\bibfnamefont {X.}~\bibnamefont {Letartre}}, \bibinfo {author}
  {\bibfnamefont {L.}~\bibnamefont {Li}}, \bibinfo {author} {\bibfnamefont
  {E.~H.}\ \bibnamefont {Linfield}}, \bibinfo {author} {\bibfnamefont {A.~G.}\
  \bibnamefont {Davies}}, \bibinfo {author} {\bibfnamefont {H.~E.}\
  \bibnamefont {Beere}}, \ and\ \bibinfo {author} {\bibfnamefont {D.~A.}\
  \bibnamefont {Ritchie}},\ }\href@noop {} {\bibfield  {journal} {\bibinfo
  {journal} {Nat Commun}\ }\textbf {\bibinfo {volume} {3}},\ \bibinfo {pages}
  {952} (\bibinfo {year} {2012})}\BibitemShut {NoStop}%
\bibitem [{sup()}]{supp}%
  \BibitemOpen
  \href@noop {} {\enquote {\bibinfo {title} {See supplementary material at ...
  for additional information including reflectance spectra at all the explored
  angles of incidence and details on the extraction of the hopfield
  coefficients.}}\ }\BibitemShut {NoStop}%
\bibitem [{\citenamefont {Whittaker}\ and\ \citenamefont
  {Culshaw}(1999)}]{Whi1}%
  \BibitemOpen
  \bibfield  {author} {\bibinfo {author} {\bibfnamefont {D.~M.}\ \bibnamefont
  {Whittaker}}\ and\ \bibinfo {author} {\bibfnamefont {I.~S.}\ \bibnamefont
  {Culshaw}},\ }\href@noop {} {\bibfield  {journal} {\bibinfo  {journal}
  {Physical Review B}\ }\textbf {\bibinfo {volume} {60}},\ \bibinfo {pages}
  {2610} (\bibinfo {year} {1999})}\BibitemShut {NoStop}%
\bibitem [{\citenamefont {Li}(1996)}]{Li1}%
  \BibitemOpen
  \bibfield  {author} {\bibinfo {author} {\bibfnamefont {L.}~\bibnamefont
  {Li}},\ }\href@noop {} {\bibfield  {journal} {\bibinfo  {journal} {J. Opt.
  Soc. Am. A}\ }\textbf {\bibinfo {volume} {13}},\ \bibinfo {pages} {1870}
  (\bibinfo {year} {1996})}\BibitemShut {NoStop}%
\bibitem [{\citenamefont {Palik}(1991)}]{Pal1}%
  \BibitemOpen
  \bibfield  {author} {\bibinfo {author} {\bibfnamefont {E.}~\bibnamefont
  {Palik}},\ }\href@noop {} {\emph {\bibinfo {title} {Handbook of Optical
  Constants of Solids}}}\ (\bibinfo  {publisher} {Academic Press},\ \bibinfo
  {address} {San Diego},\ \bibinfo {year} {1991})\BibitemShut {NoStop}%
\bibitem [{\citenamefont {Załużny}\ and\ \citenamefont
  {Nalewajko}(1999)}]{Zal1}%
  \BibitemOpen
  \bibfield  {author} {\bibinfo {author} {\bibfnamefont {M.}~\bibnamefont
  {Załużny}}\ and\ \bibinfo {author} {\bibfnamefont {C.}~\bibnamefont
  {Nalewajko}},\ }\href@noop {} {\bibfield  {journal} {\bibinfo  {journal}
  {Physical Review B}\ }\textbf {\bibinfo {volume} {59}},\ \bibinfo {pages}
  {13043} (\bibinfo {year} {1999})}\BibitemShut {NoStop}%
\bibitem [{\citenamefont {Ordal}\ \emph {et~al.}(1983)\citenamefont {Ordal},
  \citenamefont {Long}, \citenamefont {Bell}, \citenamefont {Bell},
  \citenamefont {Bell}, \citenamefont {Alexander},\ and\ \citenamefont
  {Ward}}]{Ord1}%
  \BibitemOpen
  \bibfield  {author} {\bibinfo {author} {\bibfnamefont {M.~A.}\ \bibnamefont
  {Ordal}}, \bibinfo {author} {\bibfnamefont {L.~L.}\ \bibnamefont {Long}},
  \bibinfo {author} {\bibfnamefont {R.~J.}\ \bibnamefont {Bell}}, \bibinfo
  {author} {\bibfnamefont {S.~E.}\ \bibnamefont {Bell}}, \bibinfo {author}
  {\bibfnamefont {R.~R.}\ \bibnamefont {Bell}}, \bibinfo {author}
  {\bibfnamefont {R.~W.}\ \bibnamefont {Alexander}}, \ and\ \bibinfo {author}
  {\bibfnamefont {C.~A.}\ \bibnamefont {Ward}},\ }\href@noop {} {\bibfield
  {journal} {\bibinfo  {journal} {Applied Optics}\ }\textbf {\bibinfo {volume}
  {22}},\ \bibinfo {pages} {1099} (\bibinfo {year} {1983})}\BibitemShut
  {NoStop}%
\bibitem [{\citenamefont {Hopfield}(1958)}]{Hop1}%
  \BibitemOpen
  \bibfield  {author} {\bibinfo {author} {\bibfnamefont {J.~J.}\ \bibnamefont
  {Hopfield}},\ }\href@noop {} {\bibfield  {journal} {\bibinfo  {journal}
  {Physical Review}\ }\textbf {\bibinfo {volume} {112}},\ \bibinfo {pages}
  {1555} (\bibinfo {year} {1958})}\BibitemShut {NoStop}%
\bibitem [{\citenamefont {Manceau}\ \emph {et~al.}(2013)\citenamefont
  {Manceau}, \citenamefont {Zanotto}, \citenamefont {Sagnes}, \citenamefont
  {Beaudoin},\ and\ \citenamefont {Colombelli}}]{Man7}%
  \BibitemOpen
  \bibfield  {author} {\bibinfo {author} {\bibfnamefont {J.~M.}\ \bibnamefont
  {Manceau}}, \bibinfo {author} {\bibfnamefont {S.}~\bibnamefont {Zanotto}},
  \bibinfo {author} {\bibfnamefont {I.}~\bibnamefont {Sagnes}}, \bibinfo
  {author} {\bibfnamefont {G.}~\bibnamefont {Beaudoin}}, \ and\ \bibinfo
  {author} {\bibfnamefont {R.}~\bibnamefont {Colombelli}},\ }\href@noop {}
  {\bibfield  {journal} {\bibinfo  {journal} {Applied Physics Letters}\
  }\textbf {\bibinfo {volume} {103}},\ \bibinfo {pages} {091110} (\bibinfo
  {year} {2013})}\BibitemShut {NoStop}%
\bibitem [{\citenamefont {Haus}(1984)}]{Hau1}%
  \BibitemOpen
  \bibfield  {author} {\bibinfo {author} {\bibfnamefont {H.}~\bibnamefont
  {Haus}},\ }\href@noop {} {\emph {\bibinfo {title} {Waves and fields in
  optoelectronics}}}\ (\bibinfo  {publisher} {Prentice-Hall},\ \bibinfo
  {address} {Upper Saddle River},\ \bibinfo {year} {1984})\BibitemShut
  {NoStop}%
\bibitem [{\citenamefont {Fan}, \citenamefont {Suh},\ and\ \citenamefont
  {Joannopoulos}(2003)}]{Fan1}%
  \BibitemOpen
  \bibfield  {author} {\bibinfo {author} {\bibfnamefont {S.}~\bibnamefont
  {Fan}}, \bibinfo {author} {\bibfnamefont {W.}~\bibnamefont {Suh}}, \ and\
  \bibinfo {author} {\bibfnamefont {J.~D.}\ \bibnamefont {Joannopoulos}},\
  }\href@noop {} {\bibfield  {journal} {\bibinfo  {journal} {J. Opt. Soc. Am.
  A}\ }\textbf {\bibinfo {volume} {20}},\ \bibinfo {pages} {569} (\bibinfo
  {year} {2003})}\BibitemShut {NoStop}%
\bibitem [{\citenamefont {Cai}\ \emph {et~al.}(1998)\citenamefont {Cai},
  \citenamefont {Brener}, \citenamefont {Lopata}, \citenamefont {Wynn},
  \citenamefont {Pfeiffer}, \citenamefont {Stark}, \citenamefont {Wu},
  \citenamefont {Zhang},\ and\ \citenamefont {Federici}}]{Cai1}%
  \BibitemOpen
  \bibfield  {author} {\bibinfo {author} {\bibfnamefont {Y.}~\bibnamefont
  {Cai}}, \bibinfo {author} {\bibfnamefont {I.}~\bibnamefont {Brener}},
  \bibinfo {author} {\bibfnamefont {J.}~\bibnamefont {Lopata}}, \bibinfo
  {author} {\bibfnamefont {J.}~\bibnamefont {Wynn}}, \bibinfo {author}
  {\bibfnamefont {L.}~\bibnamefont {Pfeiffer}}, \bibinfo {author}
  {\bibfnamefont {J.~B.}\ \bibnamefont {Stark}}, \bibinfo {author}
  {\bibfnamefont {Q.}~\bibnamefont {Wu}}, \bibinfo {author} {\bibfnamefont
  {X.~C.}\ \bibnamefont {Zhang}}, \ and\ \bibinfo {author} {\bibfnamefont
  {J.~F.}\ \bibnamefont {Federici}},\ }\href@noop {} {\bibfield  {journal}
  {\bibinfo  {journal} {Applied Physics Letters}\ }\textbf {\bibinfo {volume}
  {73}},\ \bibinfo {pages} {444} (\bibinfo {year} {1998})}\BibitemShut
  {NoStop}%
\bibitem [{\citenamefont {Zanotto}\ \emph {et~al.}(2014)\citenamefont
  {Zanotto}, \citenamefont {Mezzapesa}, \citenamefont {Bianco}, \citenamefont
  {Biasiol}, \citenamefont {Baldacci}, \citenamefont {Vitiello}, \citenamefont
  {Sorba}, \citenamefont {Colombelli},\ and\ \citenamefont
  {Tredicucci}}]{Zan2}%
  \BibitemOpen
  \bibfield  {author} {\bibinfo {author} {\bibfnamefont {S.}~\bibnamefont
  {Zanotto}}, \bibinfo {author} {\bibfnamefont {F.}~\bibnamefont {Mezzapesa}},
  \bibinfo {author} {\bibfnamefont {F.}~\bibnamefont {Bianco}}, \bibinfo
  {author} {\bibfnamefont {G.}~\bibnamefont {Biasiol}}, \bibinfo {author}
  {\bibfnamefont {L.}~\bibnamefont {Baldacci}}, \bibinfo {author}
  {\bibfnamefont {M.}~\bibnamefont {Vitiello}}, \bibinfo {author}
  {\bibfnamefont {L.}~\bibnamefont {Sorba}}, \bibinfo {author} {\bibfnamefont
  {R.}~\bibnamefont {Colombelli}}, \ and\ \bibinfo {author} {\bibfnamefont
  {A.}~\bibnamefont {Tredicucci}},\ }\href@noop {} {\bibfield  {journal}
  {\bibinfo  {journal} {Nature Physics}\ } (\bibinfo {year} {in press,
  2014})}\BibitemShut {NoStop}%
\bibitem [{\citenamefont {Auffèves-Garnier}\ \emph {et~al.}(2007)\citenamefont
  {Auffèves-Garnier}, \citenamefont {Simon}, \citenamefont {Gérard},\ and\
  \citenamefont {Poizat}}]{Auf1}%
  \BibitemOpen
  \bibfield  {author} {\bibinfo {author} {\bibfnamefont {A.}~\bibnamefont
  {Auffèves-Garnier}}, \bibinfo {author} {\bibfnamefont {C.}~\bibnamefont
  {Simon}}, \bibinfo {author} {\bibfnamefont {J.-M.}\ \bibnamefont {Gérard}},
  \ and\ \bibinfo {author} {\bibfnamefont {J.-P.}\ \bibnamefont {Poizat}},\
  }\href@noop {} {\bibfield  {journal} {\bibinfo  {journal} {Physical Review
  A}\ }\textbf {\bibinfo {volume} {75}},\ \bibinfo {pages} {053823} (\bibinfo
  {year} {2007})}\BibitemShut {NoStop}%
\end{thebibliography}
\end{document}